\documentstyle[prd,aps,preprint,tighten,epsfig]{revtex}

\begin{document}

\draft

\title{Hierarchy and Up-Down Parallelism of Quark Mass Matrices}
\author{{\bf Jian-wei Mei} ~ and ~ {\bf Zhi-zhong Xing}}
\address{Institute of High Energy Physics, 
Chinese Academy of Sciences, \\
P.O. Box 918 (4), Beijing 100039, China 
\footnote{Mailing address} \\
({\it Electronic address: xingzz@mail.ihep.ac.cn}) }
\maketitle

\begin{abstract}
In view of the quark mass hierarchy and in the assumption of
the up-down parallelism, we derive two phenomenologically-favored 
patterns of Hermitian quark mass matrices from the quark 
flavor mixing matrix. We compare one of them with two existing 
{\it Ans$\it\ddot{a}$tze} proposed by Rosner and 
Worah and by Roberts {\it et al}, and find that only the
latter is consistent with the present experimental data.
\end{abstract}


\newpage

\section{Introduction}

The masses and flavor mixing angles of six quarks are free parameters
in the standard electroweak model, but their values can directly or
indirectly be determined from a number of experiments \cite{PDG}. 
The fact that three quark masses in 
each (up or down) sector perform a strong hierarchy is a big puzzle 
to particle physicists, so is the hierarchy of three quark mixing 
angles. Theoretical attempts to solve the puzzle, e.g., those starting 
from supersymmetric grand unification 
theories or from superstring theories \cite{Ibanez}, are encouraging 
but have not proved to be very successful. Phenomenologically, the
common approach is to find out simple textures of quark mass matrices, 
from which a bridge between the hierarchy of quark mixing angles 
and that of quark masses can naturally be established \cite{Review}. 
The flavor symmetries hidden in such textures might finally provide us 
with useful hints towards the underlying dynamics responsible for 
the generation of quark masses and the origin of $CP$ violation. 

The Cabibbo-Kobayashi-Maskawa (CKM) quark mixing matrix $V$ arises 
from the mismatch between the diagonalization of the up-type quark 
mass matrix $M_{\rm U}$ and that of the down-type quark mass matrix 
$M_{\rm D}$ \cite{CKM}. 
Without loss of generality, one may arrange $M_{\rm U}$ 
and $M_{\rm D}$ to be Hermitian in the framework of the standard model 
or its extensions which have no flavor-changing right-handed 
currents \cite{FX99}. There exist two reciprocal ways to explore 
possible relations between the hierarchy of $V$ and that
of $M_{\rm U}$ and $M_{\rm D}$:

(a) Taking account of a specific texture of the Hermitian quark
mass matrix $M_{\rm U}$ or $M_{\rm D}$, which might result from
a new flavor symmetry and its explicit breaking or rely on some 
plausible theoretical arguments, one may do the unitary transformation
\begin{eqnarray}
O^\dagger_{\rm U} M_{\rm U} O_{\rm U} & = &
\left ( \matrix{
\delta_u	& 0	& 0 \cr
0 	& \delta_c	& 0 \cr
0	& 0	& \delta_t \cr} \right ) \; ,
\nonumber \\
O^\dagger_{\rm D} M_{\rm D} O_{\rm D} & = &
\left ( \matrix{
\delta_d	& 0	& 0 \cr
0 	& \delta_s	& 0 \cr
0	& 0	& \delta_b \cr} \right ) \; ,
%	(1)
\end{eqnarray}
where $\delta_i$ stands for the mass eigenvalue of the $i$-quark.
Then the CKM matrix $V$ is given as 
$V = O^\dagger_{\rm U} O_{\rm D}$, from which the quark mixing
angles can be expressed in terms of the ratios of quark masses
(and other unfixed parameters of $M_{\rm U}$ and $M_{\rm D}$).
This {\it normal} approach has extensively been discussed in the 
literature \cite{Review}.

(b) Taking account of the experimentally-allowed pattern of $V$ 
and making some proper assumptions to decompose $V$ into 
$V = O^\dagger_{\rm U} O_{\rm D}$ (e.g., the principle of 
``naturalness'' disfavors any severe fine-tuning of or delicate 
cancellations between relevant parameters in the unitary 
transformation matrices
$O_{\rm U}$ and $O_{\rm D}$), one may construct the Hermitian
quark mass matrices $M_{\rm U}$ and $M_{\rm D}$ through
\begin{eqnarray}
M_{\rm U} & = &
O_{\rm U} \left ( \matrix{
\delta_u	& 0	& 0 \cr
0 	& \delta_c	& 0 \cr
0	& 0	& \delta_t \cr} \right ) O^\dagger_{\rm U} \; ,
\nonumber \\
M_{\rm D} & = &
O_{\rm D} \left ( \matrix{
\delta_d	& 0	& 0 \cr
0 	& \delta_s	& 0 \cr
0	& 0	& \delta_b \cr} \right ) O^\dagger_{\rm D} \; .
%	(2)
\end{eqnarray}
This {\it reverse} approach has also attracted some
attention in the literature \cite{RRR}.

In this Brief Report, we present a further study of the 
relationship between the CKM matrix and the quark mass matrices 
by following approach (b). Our main goal is to derive the 
phenomenologically-favored textures of Hermitian quark mass 
matrices from the CKM matrix. We are guided by 
the hierarchy of quark masses, the hierarchy of the CKM matrix 
elements and the afore-mentioned principle of naturalness. 
The only assumption that we need to make is the structural 
parallelism between  
up- and down-quark mass matrices. Such an empirical assumption 
should essentially be true, if two quark sectors are governed by 
the same dynamics. We arrive at two explicit textures of quark
mass matrices, which were not obtained in Ref. \cite{RRR}. One 
of them is particularly interesting, as its form is 
somehow similar to the {\it Ans$\it\ddot{a}$tze} 
proposed by Rosner and Worah \cite{Rosner} and by 
Roberts {\it et al} \cite{Roberts}. A careful comparison
shows that our pattern of quark mass matrices is compatible 
with the ansatz advocated in Ref. \cite{Roberts}, while the 
Rosner-Worah pattern is not favored by the present experimental 
data.

\section{Hierarchy and up-down parallelism}

The hierarchy of quark flavor mixing can clearly be seen from 
the Wolfenstein parametrization of the CKM matrix $V$ \cite{Wol}:
\begin{equation}
V \; \approx \; \left ( \matrix{
1 - \frac{1}{2} \lambda^2 - \frac{1}{8} \lambda^4 & \lambda 
& p\lambda^3 \cr
- \lambda & 
1 - \frac{1}{2} \lambda^2 - \frac{1}{8} \left (1 + 4A^2 \right )
\lambda^4 & A\lambda^2 \cr
q\lambda^3 & 
-A\lambda^2 + \frac{1}{2} \left (2q - A \right ) \lambda^4 & 
1 - \frac{1}{2} A^2 \lambda^4 \cr} \right ) \; ,
%        (3)
\end{equation}
where $p\equiv A(\rho -i\eta)$, $q\equiv A (1-\rho-i\eta)$, and
the terms of or below ${\cal O}(\lambda^5)$ have been neglected. 
Current experimental data yield $\lambda \approx 0.22$, 
$A \approx 0.83$, $\rho \approx 0.17$ and 
$\eta \approx 0.36$ \cite{Buras}. Thus we have
\begin{eqnarray} 
|p| & = & A\sqrt{\rho^2 + \eta^2} \; \approx \; 0.33 \; ,
\nonumber \\
|q| & = & A\sqrt{(1-\rho)^2 + \eta^2} \; \approx \; 0.75 \; .
%        (4)
\end{eqnarray}
We find that $|p|/|q| \approx 2\lambda$ holds and 
$|V_{ub}| = |p|\lambda^3$ is actually of ${\cal O}(\lambda^4)$.

To construct the Hermitian quark mass matrices $M_{\rm U}$ and 
$M_{\rm D}$ from $V$ through Eq. (2), the key point is to 
determine the unitary transformation matrices $O_{\rm U}$ and 
$O_{\rm D}$.
As the physical quark masses $(m_u, m_c, m_t)$ or $(m_d, m_s, m_b)$
approximately perform a geometrical hierarchy at a common 
energy scale (e.g., the electroweak scale $\mu = M_Z$ \cite{Review}),
\begin{eqnarray}
\frac{m_u}{m_c} & \sim & \frac{m_c}{m_t} \; \sim \; \zeta^2 \; ,
\nonumber \\
\frac{m_d}{m_s} & \sim & \frac{m_s}{m_b} \; \sim \; \lambda^2 \; ,
%        (5)
\end{eqnarray}
where $\zeta$ is of ${\cal O}(\lambda^2)$ and can be defined as
$\zeta \equiv \kappa \lambda^2$ with $\kappa \sim {\cal O}(1)$,
we are allowed to expand the matrix elements of $M_{\rm U}$ or
$M_{\rm D}$ in powers of $\zeta$ or $\lambda$.
Provided two quark sectors are governed by the same dynamics,
$M_{\rm U}$ and $M_{\rm D}$ are then expected to have approximately
parallel structures characterized respectively by the perturbative
parameters $\zeta$ and $\lambda$. This up-down parallelism would 
be exact, if the geometrical relations in Eq. (5) held exactly
(i.e., $m_u/m_c = m_c/m_t = \zeta^2$ and
$m_d/m_s = m_s/m_b = \lambda^2$). Note that the parallelism
between $M_{\rm U}$ and $M_{\rm D}$ implies the parallelism
between $O_{\rm U}$ and $O_{\rm D}$, whose textures can respectively 
be expanded in terms of $\zeta$ and $\lambda$. Since the CKM matrix
$V = O^\dagger_{\rm U} O_{\rm D}$ deviates from the unity matrix only
at the ${\cal O}(\lambda)$ level, both $O_{\rm U}$ and $O_{\rm D}$ 
must be close to the unity matrix in the spirit of naturalness. 
It is therefore instructive to expand $O_{\rm U}$ and $O_{\rm D}$ 
as follows:
\begin{eqnarray}
O_{\rm U} & = & I + \sum^{\infty}_{n=1} 
\left ( Z^{\rm U}_n \zeta^n \right ) \; ,
\nonumber \\
O_{\rm D} & = & I + \sum^{\infty}_{n=1} 
\left ( Z^{\rm D}_n \lambda^n \right ) \; ,
%        (6)
\end{eqnarray}
where $I$ stands for the unity matrix, and $Z^{\rm U}_n$ or
$Z^{\rm D}_n$ (for $n=1,2,3,\cdot\cdot\cdot$) denotes the coefficient
matrix of $\zeta^n$ or $\lambda^n$. While 
$Z^{\rm U}_n \sim Z^{\rm D}_n$ should in general hold, the simplest
possibility is $Z^{\rm U}_n = Z^{\rm D}_n$ ($\equiv Z_n$), equivalent 
to the exact parallelism between $O_{\rm U}$ and $O_{\rm D}$. 
In this especially interesting case, the expressions of 
$Z_1$, $Z_2$, $Z_3$ and $Z_4$ can concretely be determined from 
Eqs. (3) and (6):
\begin{eqnarray}
Z_1 & = & \left ( \matrix{
0	& ~~ 1 ~	& ~ 0 \cr
-1	& ~~ 0 ~	& ~ 0 \cr
0	& ~~ 0 ~	& ~ 0 \cr} \right ) \; ,
\nonumber \\
Z_2 & = & \left ( \matrix{
-\frac{1}{2}	& \kappa	& 0 \cr
-\kappa	& -\frac{1}{2}	& A \cr
0	& -A	& 0 \cr} \right ) \; ,
\nonumber \\
Z_3 & = & \left ( \matrix{
-\kappa	& 0	& ~ p \cr
0	& -\kappa	& ~ 0 \cr
q	& 0	& ~ 0 \cr} \right ) \; ,
\nonumber \\
Z_4 & = & \left ( \matrix{
-\frac{1}{8} \left (1 + 4\kappa^2 \right )	&
-\frac{1}{2} \kappa \left (1 + 2\kappa^2 \right )	&
A\kappa \cr
\frac{1}{2} \kappa \left (1 - 2\kappa^2 \right )	& 
-\frac{1}{8} \left ( 1 + 4\kappa^2 + 4A^2\right ) & 
A\kappa^2 \cr
0	& 
-A\kappa^2 + \frac{1}{2} \left (A - 2p^* \right )	& 
-\frac{1}{2} A^2 \cr} \right ) \; .
%	(7)
\end{eqnarray}
We see that the complex $CP$-violating phase enters $O_{\rm U}$ 
or $O_{\rm D}$ at the level of ${\cal O}(\zeta^3)$ or
${\cal O}(\lambda^3)$. As a straightfoward consequence of 
$\zeta \sim {\cal O}(\lambda^2)$, the contribution of $O_{\rm D}$ 
to $V$ is dominant over that of $O_{\rm U}$ to $V$. 
  
With the help of Eqs. (6) and (7), we are now able to derive
the hierarchical textures of $M_{\rm U}$ and $M_{\rm D}$ from 
Eq. (2). Note that the quark mass eigenvalues 
$(\delta_u, \delta_c, \delta_t)$ and
$(\delta_d, \delta_s, \delta_b)$ in Eq. (2) may be either 
positive or negative. Without loss of generality, we take 
$\delta_t = m_t$ and $\delta_b = m_b$. The other four mass 
eigenvalues can be expressed, in view of Eq. (5), as follows:
\begin{eqnarray}
\delta_u & = & r_u \zeta^4 m_t \; ,
\nonumber \\
\delta_c & = & r_c \zeta^2 m_t \; ;
%        (8)
\end{eqnarray}
and
\begin{eqnarray}
\delta_d & = & r_d \lambda^4 m_b \; ,
\nonumber \\
\delta_s & = & r_s \lambda^2 m_b \; ,
%        (9)
\end{eqnarray}
where $|r_u| \sim |r_c| \sim |r_d| \sim |r_s| \sim {\cal O}(1)$
holds. Then we arrive at the results of $M_{\rm U}$ and 
$M_{\rm D}$: 
\begin{eqnarray}
M_{\rm U} & \approx & m_t \left ( \matrix{
(r_u + r_c) \zeta^4
& ~ \tilde{r}_c \zeta^3
& p^{~}_\zeta \zeta^3 \cr
\tilde{r}_c \zeta^3
& ~ r_c \zeta^2
& A \zeta^2 \cr
p^*_\zeta \zeta^3
& ~ A \zeta^2
& 1 \cr} \right ) \; ,
\nonumber \\
M_{\rm D} & \approx & m_b \left ( \matrix{
(r_d + r_s) \lambda^4
& \tilde{r}_s \lambda^3
& p^{~}_\lambda \lambda^3 \cr
\tilde{r}_s \lambda^3
& r_s \lambda^2
& A \lambda^2 \cr
p^*_\lambda \lambda^3
& A \lambda^2
& 1 \cr} \right ) \; ,
%        (10)
\end{eqnarray}
where 
\begin{eqnarray}
\tilde{r}_c & \equiv & r_c \left (1 + \kappa \zeta \right ) \; ,
\nonumber \\
\tilde{r}_s & \equiv & r_s \left (1 + \kappa \lambda \right ) \; ,
%	(11)
\end{eqnarray}
and
\begin{eqnarray}
p^{~}_\zeta & \equiv & p + \kappa A \zeta \; ,
\nonumber \\
p^{~}_\lambda & \equiv & p + \kappa A \lambda \; .
%	(12)
\end{eqnarray}
We see that the parallelism between $M_{\rm U}$ and $M_{\rm D}$
in Eq. (10) is not exact (e.g., 
$p^{~}_\zeta \approx p \neq p^{~}_\lambda$), even though the 
exact parallelism between $O_{\rm U}$ and $O_{\rm D}$ has been 
assumed in our calculations.

Note that the sign uncertainties of quark mass eigenvalues may
lead to two distinct textures of $M_{\rm U}$ and $M_{\rm D}$.
If $r_u$ (or $r_d$) and $r_c$ (or $r_s$) have the same sign,
the (1,1) element of $M_{\rm U}$ (or $M_{\rm D}$) amounts 
approximately to $2r_c \zeta^4$ (or $2r_s \lambda^4$). In this
case, we obtain
\begin{eqnarray}
M_{\rm U} & \approx & m_t \left ( \matrix{
2r_c \zeta^4
& ~ \tilde{r}_c \zeta^3
& p^{~}_\zeta \zeta^3 \cr
\tilde{r}_c \zeta^3
& ~ r_c \zeta^2
& A \zeta^2 \cr
p^*_\zeta \zeta^3
& ~ A \zeta^2
& 1 \cr} \right ) \; ,
\nonumber \\
M_{\rm D} & \approx & m_b \left ( \matrix{
2 r_s \lambda^4
& \tilde{r}_s \lambda^3
& p^{~}_\lambda \lambda^3 \cr
\tilde{r}_s \lambda^3
& r_s \lambda^2
& A \lambda^2 \cr
p^*_\lambda \lambda^3
& A \lambda^2
& 1 \cr} \right ) \; .
%        (13)
\end{eqnarray}
If the signs of $r_u$ (or $r_d$) and $r_c$ (or $r_s$) are 
opposite to each other, however, a very significant cancellation 
must appear between them 
%%%%%%%%%%%%%%%%%%%%%%%%%%%%%
\footnote{This cancellation does not involve two different quark 
sectors, thus it has no conflict with the principle of 
naturalness mentioned before.}.
%%%%%%%%%%%%%%%%%%%%%%%%%%%%% 
In this case, the (1,1) elements of
$M_{\rm U}$ and $M_{\rm D}$ are expected to be of or below 
${\cal O}(\zeta^5)$ and ${\cal O}(\lambda^5)$, respectively.
Then we are led to a somehow simpler texture of quark mass
matrices:
\begin{eqnarray}
M_{\rm U} & \approx & m_t \left ( \matrix{
0
& ~ \tilde{r}_c \zeta^3
& p^{~}_\zeta \zeta^3 \cr
\tilde{r}_c \zeta^3
& ~ r_c \zeta^2
& A \zeta^2 \cr
p^*_\zeta \zeta^3
& ~ A \zeta^2
& 1 \cr} \right ) \; ,
\nonumber \\
M_{\rm D} & \approx & m_b \left ( \matrix{
0
& \tilde{r}_s \lambda^3
& p^{~}_\lambda \lambda^3 \cr
\tilde{r}_s \lambda^3
& r_s \lambda^2
& A \lambda^2 \cr
p^*_\lambda \lambda^3
& A \lambda^2
& 1 \cr} \right ) \; .
%        (14)
\end{eqnarray}
The textures of Hermitian quark mass matrices in Eqs. (13) and
(14) result naturally from the quark mass hierarchy and the 
up-down parallelism. Therefore they can be regarded as two
promising candidates for the ``true'' quark mass matrices 
in an underlying effective theory of quark mass generation at 
low energies. 

\section{Comparison with two existing Ans$\bf\ddot{a}$tze}

Now let us compare the texture of quark mass matrices obtained
in Eq. (14) with two existing {\it Ans$\it\ddot{a}$tze} proposed 
by Rosner and Worah (RW) in Ref. \cite{Rosner} and by 
Roberts, Romanino, Ross and Velasco-Sevilla (RRRV)
in Ref. \cite{Roberts}. Note that quark mass 
matrices of both RW and RRRV forms are symmetric instead of 
Hermitian. Their consequences on quark flavor mixing and $CP$ 
violation are essentially unchanged, however, if the Hermiticity is 
imposed on them \cite{Roberts}. For this reason, we consider 
the Hermitian versions of the RW and RRRV  
{\it Ans$\it\ddot{a}$tze} and compare them with Eq. (14).

The RW ansatz of quark mass matrices is based on a 
composite model of spin-1/2 particles \cite{Rosner} and its
Hermitian form can be written as
\begin{eqnarray}
{\cal M}_{\rm U} & = & \left ( \matrix{
0 & \sqrt{2} \alpha^{~}_{\rm U} & \alpha^{~}_{\rm U} \cr
\sqrt{2} \alpha^*_{\rm U} & 
\beta_{\rm U} & \sqrt{2} \beta_{\rm U} \cr
\alpha^*_{\rm U} & \sqrt{2} \beta_{\rm U} & \gamma^{~}_{\rm U} \cr}
\right ) \; ,
\nonumber \\
{\cal M}_{\rm D} & = & \left ( \matrix{
0 & \sqrt{2} \alpha^{~}_{\rm D} & \alpha^{~}_{\rm D} \cr
\sqrt{2} \alpha^*_{\rm D} & 
\beta_{\rm D} & \sqrt{2} \beta_{\rm D} \cr
\alpha^*_{\rm D} & \sqrt{2} \beta_{\rm D} & \gamma^{~}_{\rm D} \cr}
\right ) \; ,
%        (15)
\end{eqnarray}
where $\alpha^{~}_{\rm U}$ and $\alpha^{~}_{\rm D}$ are complex 
parameters so as to accommodate the observed effects of 
$CP$ violation in the quark sector \cite{PDG}.
Comparing between Eqs. (14) and (15), we find that the 
latter could basically be reproduced from the former, 
if the following conditions were satisfied:
\begin{eqnarray}
\frac{|\tilde{r}_c|}{|p^{~}_\zeta|} & \approx & \frac{A}{|r_c|} \; 
\approx \; \sqrt{2} \; ,
\nonumber \\
\frac{|\tilde{r}_s|}{|p^{~}_\lambda|} & \approx & \frac{A}{|r_s|} \; 
\approx \; \sqrt{2} \; .
%        (16)
\end{eqnarray}
In view of Eqs. (8) and (9) as well as Eqs. (11) and (12), we
immediately realize that the relations in Eq. (16) are
impossible to hold. Hence the RW texture is actually 
{\it incompatible} with ours given in Eq. (14). Does this 
incompatibility imply the disagreement between the RW 
ansatz and current experimental data? We find that the answer is 
affirmative. Since the (1,2) and (1,3) elements of ${\cal M}_{\rm U}$ 
or ${\cal M}_{\rm D}$ in Eq. (15) are comparable in magnitude, we are 
led to the prediction 
\begin{equation}
|V_{td}| \; \approx \; |V_{ub}| \; \sim \; 
\frac{1}{\sqrt{2}} \lambda^3 \;
%        (17)
\end{equation}
in the leading-order
approximation. Such a result is obviously inconsistent with the
present experimental data, which require 
$|V_{ub}| \sim {\cal O}(\lambda^4)$ 
and $|V_{ub}/V_{td}| \approx |p|/|q| \approx 2\lambda$. Thus
the RW texture of quark mass matrices, no matter whether it is
Hermitian or symmetric, is no longer favored in phenomenology.

The Hermitian RRRV ansatz of quark mass matrices \cite{Roberts} takes 
the form
%%%%%%%%%%%%%%%%%%%%%%%%%%%%%%%%%%
\footnote{This texture is indeed similar to the one discussed by
Branco {\it et al} in Ref. \cite{Branco}.}
%%%%%%%%%%%%%%%%%%%%%%%%%%%%%%%%%%
\begin{eqnarray}
{\cal M}_{\rm U} & = & m_t \left ( \matrix{
0	& b_{\rm U} \epsilon^3_{\rm U} 	
& c^{~}_{\rm U} \epsilon^4_{\rm U} \cr
b^*_{\rm U} \epsilon^3_{\rm U}	& \epsilon^2_{\rm U}
& a^{~}_{\rm U} \epsilon^2_{\rm U} \cr
c^*_{\rm U} \epsilon^4_{\rm U}	& a^*_{\rm U} \epsilon^2_{\rm U}
& 1 \cr} \right ) \; ,
\nonumber \\
{\cal M}_{\rm D} & = & m_b \left ( \matrix{
0	& b_{\rm D} \epsilon^3_{\rm D} 	
& c^{~}_{\rm D} \epsilon^4_{\rm D} \cr
b^*_{\rm D} \epsilon^3_{\rm D}	& \epsilon^2_{\rm D}
& a^{~}_{\rm D} \epsilon^2_{\rm D} \cr
c^*_{\rm D} \epsilon^4_{\rm D}	& a^*_{\rm D} \epsilon^2_{\rm D}
& 1 \cr} \right ) \; ,
%	(18)
\end{eqnarray}
where $\epsilon^{~}_{\rm U} \approx \sqrt{m_c/m_t}$ and 
$\epsilon^{~}_{\rm D} \approx \sqrt{m_s/m_b}$ stand 
respectively for the expansion parameters of up- and down-quark
sectors, and the remaining parameters 
$(a^{~}_{\rm U}, b_{\rm U}, c^{~}_{\rm U})$ and 
$(a^{~}_{\rm D}, b_{\rm D}, c^{~}_{\rm D})$ are all
of ${\cal O}(1)$ in magnitude. From Eqs. (8) and (9), 
we obtain the relation between 
$\epsilon^{~}_{\rm U}$ (or $\epsilon^{~}_{\rm D}$) and $\zeta$ 
(or $\lambda$) as follows:
\begin{eqnarray}
\epsilon^{~}_{\rm U} & \approx & \zeta \sqrt{|r_c|} \;\; ,
\nonumber \\
\epsilon^{~}_{\rm D} & \approx & \lambda \sqrt{|r_s|} \;\; .
%	(19)
\end{eqnarray}
In addition, we find
\begin{eqnarray}
|a^{~}_{\rm U}| & \approx & \frac{A}{|r_c|} \; ,
\nonumber \\
|b_{\rm U}| & \approx & \frac{1 + \kappa \zeta}{\sqrt{|r_c|}} \; ,
\nonumber \\
|c^{~}_{\rm U}| & \approx & \frac{|p^{~}_\zeta|}
{\zeta |r_c|^2} \; ;
%	(20)
\end{eqnarray}
and
\begin{eqnarray}
|a^{~}_{\rm D}| & \approx & \frac{A}{|r_s|} \; ,
\nonumber \\
|b_{\rm D}| & \approx & \frac{1 + \kappa \lambda}{\sqrt{|r_s|}} \; ,
\nonumber \\
|c^{~}_{\rm D}| & \approx & \frac{|p^{~}_\lambda|}
{\lambda |r_s|^2} \; .
%	(21)
\end{eqnarray}
It is obvious that the moduli of 
$(a^{~}_{\rm U}, b_{\rm U}, c^{~}_{\rm U})$ and
$(a^{~}_{\rm D}, b_{\rm D}, c^{~}_{\rm D})$ are all of 
${\cal O}(1)$, consistent with the assumption made in 
Ref. \cite{Roberts}
%%%%%%%%%%%%%%%%%%%%%%%%%%%
\footnote{Note that 
$|c^{~}_{\rm U}| \sim \lambda/\zeta \sim 1/\lambda$ may 
be around 4.5, but it has little impact on the global fit of 
the RRRV ansatz to current experimental data \cite{Roberts}.}.
%%%%%%%%%%%%%%%%%%%%%%%%%%%
Hence the RRRV ansatz is {\it compatible} with our texture of 
quark mass matrices given in Eq. (14) and favored by current
experimental data on quark flavor mixing.

It is worth remarking that a simplification of the RRRV
ansatz to the following four-zero texture 
\begin{eqnarray}
{\cal M}_{\rm U} & = & m_t \left ( \matrix{
0	& b_{\rm U} \epsilon^3_{\rm U} 	
& 0 \cr
b^*_{\rm U} \epsilon^3_{\rm U}	& \epsilon^2_{\rm U}
& a^{~}_{\rm U} \epsilon^2_{\rm U} \cr
0	& a^*_{\rm U} \epsilon^2_{\rm U}
& 1 \cr} \right ) \; ,
\nonumber \\
{\cal M}_{\rm D} & = & m_b \left ( \matrix{
0	& b_{\rm D} \epsilon^3_{\rm D} 	
& 0 \cr
b^*_{\rm D} \epsilon^3_{\rm D}	& \epsilon^2_{\rm D}
& a^{~}_{\rm D} \epsilon^2_{\rm D} \cr
0	& a^*_{\rm D} \epsilon^2_{\rm D}
& 1 \cr} \right ) \; 
%	(22)
\end{eqnarray}
would give rise to the prediction
$|V_{ub}|/|V_{cb}| \approx \sqrt{m_u/m_c} \leq 0.06$ \cite{FX95}
for reasonable values of $m_u$ and $m_c$ \cite{Leutwyler},
which is difficult to agree with the present experimental result
$|V_{ub}/V_{cb}|_{\rm ex} \approx 0.09$ \cite{PDG}. Such a
discrepancy implies that the interesting four-zero pattern of 
quark mass matrices in Eq. (22) might no longer be favored
%%%%%%%%%%%%%%%%%%%%%%%%%%%%%
\footnote{Note, however, that the general four-zero texture of 
Hermitian quark mass matrices can still be in good agreement
with current experimental data. See Ref. \cite{FX03} for 
detailed discussions.}.
%%%%%%%%%%%%%%%%%%%%%%%%%%%%%

Comparing Eq. (18) with Eqs. (15) and (22), we see that 
the (1,3) elements of ${\cal M}_{\rm U}$ and ${\cal M}_{\rm D}$ 
should be neither larger than or comparable 
with their neighboring (1,2) elements nor vanishing or
negligibly small. This observation is particularly true
for $M_{\rm D}$, which contributes dominantly to the CKM
matrix $V$. 

\section{Summary}

We have derived two phenomenologically-favored textures of 
Hermitian quark mass matrices from the CKM matrix. Our
starting points of view include the hierarchy of quark 
masses, the hierarchy of the CKM matrix elements and the 
principle of naturalness. The main assumption that we 
have made is the structural parallelism between up- and 
down-quark mass matrices, which is expected to be true if 
two quark sectors are governed by the same dynamics. 

We have compared one of the obtained textures with two
existing {\it Ans$\it\ddot{a}$tze} of quark mass matrices,
proposed by Rosner and Worah and by Roberts {\it et al}. It 
turns out that the Rosner-Worah ansatz is no more favored
in phenomenology, while the ansatz of Roberts {\it et al}
is in good agreement with current experimental data.

We hope that our results for the structures of quark mass
matrices may serve as useful guides to model building,
from which some deeper understanding of quark masses,
flavor mixing and $CP$ violation can finally be achieved.

\acknowledgments{This work was supported in part by the
National Natural Science Foundation of China.}

\end{document}